\documentclass{iopart}

 \expandafter\let\csname equation*\endcsname\relax
 \expandafter\let\csname endequation*\endcsname\relax

\usepackage{amsmath}
\usepackage{amssymb}
\usepackage{graphicx}
\usepackage{esint}

\providecommand{\nbraket}[1]{\left\langle#1\right\rangle}

\let \Re \relax
\let \Im \relax
\DeclareMathOperator{\Re}{Re}
\DeclareMathOperator{\Im}{Im}

\def \beq {\begin{equation}}
\def \edq {\end{equation}}
\def \bes {\begin{subequations}}
\def \eds {\end{subequations}}
\def \beqn {\begin{equation*}}
\def \edqn {\end{equation*}}

\def \dag {\dagger}

\def \up {\uparrow}
\def \down {\downarrow}

\def \sm {\sigma}

\def \veps {\varepsilon}

\def \calh {{\cal{H}}}

\def \calz {{\cal{Z}}}

\def \calg {{\cal{G}}}

\def \calt {{\cal{T}}}

\def \cals {{\cal{S}}}
\def \cali {{\cal{I}}}

\def \tGamma {\widetilde{\Gamma}}

\def \hc {\text{H.c.}}

\begin{document}

\title{Orbital caloritronic transport in strongly interacting quantum dots}

\author{Jong Soo Lim$^1$, Rosa L\'opez$^{1,2}$ and David S\'anchez$^{1,2}$}

\address{$^1$Institut de F\'{\i}sica Interdisciplin\`aria i Sistemes Complexos IFISC (UIB-CSIC),
E-07122 Palma de Mallorca, Spain\\
$ˆ{^2}$Departament de F\'{\i}sica, Universitat de les Illes Balears, E-07122 Palma de Mallorca, Spain}
\ead{lim.jongsoo@gmail.com}



\begin{abstract}
We discuss out-of-equilibrium population imbalances between different orbital states
due to applied thermal gradients. This purely thermoelectric orbital effect appears quite generically
in nanostructures with a pseudospin degree of freedom. We establish an orbital Seebeck coefficient
that characterizes the induced orbital bias in response to a temperature difference between
reservoirs coupled to a quantum conductor. We analyze a two-terminal
strongly interacting quantum dot with two orbital states and
find that the orbital thermopower acts as an excellent tool to describe
the transition between SU(4) and SU(2) Kondo physics. Our conclusions
are reinforced from a detailed comparison with the charge thermopower
using numerical renormalization group calculations.
\end{abstract}

\submitto{\NJP}

\maketitle

\section{Introduction}
 
The discovery of the spin Seebeck effect \cite{uch08}
has ignited research in spin caloritronics, a field
where the focus is put on the generation of spin-polarized
electric currents by applying thermal gradients \cite{bau10}.
Nonequilibrium spin accumulations can thus be generated
in response to a temperature difference across a junction
even when the charge current vanishes \cite{joh87}.
 
In addition to electronic spin, many nanostructures offer
the possibility of an extra degree of freedom---the orbital
quantum number. This property naturally arises in carbon nanotubes
as a result of the two ways (clockwise and anticlockwise)
that electrons possess to move around the tube axis \cite{min04}
or can be artificially realized in double quantum dot structures
since each individual dot state can be viewed as the possible
outcome of two-level pseudospin (orbital) measurements \cite{van02}.

We here put forward the idea of generating different orbital populations
using temperature gradients (orbital caloritronics). 
We thus define the orbital or pseudospin
Seebeck coefficient which measures the orbital bias voltage
generated across a mesoscopic conductor under the conditions
of vanishing charge and orbital currents. Remarkably, we find
that the orbital thermopower acts as an efficient probe to characterize
pseudospin-driven quantum transitions in quantum dots and carbon nanotubes.

The orbital degree of freedom plays an essential role in the formation
of highly symmetric Kondo states. While the conventional SU(2)
Kondo resonance arises, at low temperature $T$, from the many-body exchange interaction
between a localized spin $1\over 2$ (the quantum impurity)
and conduction band electrons (the Fermi sea), the SU(4) Kondo physics
occurs because the entangled spin and orbital degrees of freedom
form a hyperspin with higher dimensionality that undergoes simultaneous
flip processes both in the spin and the orbital sectors \cite{bor03}. Hence, the transition
between SU(2) and SU(4) Kondo effects involves two strongly correlated
states with rather different temperature scales (the Kondo temperatures
$T_K^{\rm SU(2)}$ and $T_K^{\rm SU(4)}$ that typicially fulfill $T_K^{\rm SU(4)}\gg T_K^{\rm SU(2)}$).
Such transition has been investigated both experimentally \cite{jar05,sas04,hol04}
and theoretically \cite{bor03,leh03,lop05,cho05,gal05,lim06,mak07,sil07}.
In particular, an applied magnetic field in nanotubes couples differently
to the spin and orbital quantum numbers \cite{min04}, lifting the degeneracy
and allowing for a tunable conversion from SU(4) Kondo physics
to a purely spin or orbital Kondo effect \cite{jar05}. On the other hand,
pseudospin resolved transport has been achieved very recently
in double quantum dots \cite{ama13}.

Thermoelectric properties of SU(2) Kondo impurities show not only
clear changes depending on the ratio $T/T_K$ \cite{cos10}
but also deviations from the semiclassical Mott formula \cite{sch05}.
These are important features for strongly interacting quantum dots that
might potentially work as nanoscale thermoelectric coolers
or heat-to-electricity converters. When the pseudospin degree of freedom
is created by charged states in negative charging energy quantum dots,
the Seebeck coefficient can be substantially enlarged \cite{and12}. Moreover,
pure spin currents can be thermally generated from an artificial Kondo impurity
coupled to ferromagnetic leads \cite{swi09} or in the presence of magnetic fields \cite{rej12}.
Recently, the Seebeck coefficient has been proposed as a sensitive
probe of the transition between SU(2) and SU(4) Kondo states \cite{rou12}.
It is thus natural to ask whether the generalization to orbital
thermopower can provide additional insight on that transition.
Below, we demonstrate that the orbital Seebeck coefficient
shows a characteristic minimum that signals the crossover
from one Kondo state to another. Therefore, investigation of
orbital thermoelectric effects is interesting from both viewpoints---the practical
motivation that leads to the generation
of orbital polarizations and the fundamental study
of orbital driven phase transitions and crossovers.


\section{Orbital and charge Seebeck coefficients}

We consider a generic mesoscopic conductor with interacting electrons
and coupled to left ($L$) and right ($R$) leads. Let $\nu = \pm$ be the
orbital index that labels the two orbital states present both in the sample
and in the leads. For completeness, we also take into account the spin index
$\sm = \up/\down$, although in what follows we will assume spin degeneracy
in order to focus on orbital effects only. The exact formula for the current at channel $\nu$ reads \cite{mei92,sun02}:
\beq
I_{\nu} 
= \frac{e}{h} \sum_{\sm}\int d\veps\, \left(f_{L\nu}(\veps)-f_{R\nu}(\veps)\right) \calt_{\nu\sm}(\veps) \,,
\label{eq:IMGO}
\edq
where the generalized transmission function is
\beq
\calt_{\nu\sm}(\veps) = \frac{4\pi\Gamma_L\Gamma_R}{\Gamma_L+\Gamma_R} A_{\nu\sm}(\veps)\,,
\edq
in terms of level broadenings $\Gamma_{\alpha} = \pi\sum_k |t_{\alpha}|^2 \delta(\veps-\veps_{\alpha k})$
with $t_{\alpha}$ the tunnel amplitude from lead $\alpha =L,R$. The total linewidth is then
$\Gamma = \sum_{\alpha}\Gamma_{\alpha}$.
The dot spectral weight $A_{\nu\sm}(\veps)$ in an orbital $\nu$ with spin $\sm$ is obtained from the retarded dot Green's function $\calg_{\nu\sm,\nu\sm}^r(\veps)$ by
$A_{\nu\sm}(\veps) = - \Im \calg_{\nu\sm,\nu\sm}^r(\veps)/\pi$.

In Eq.~\eqref{eq:IMGO}, the leads are Fermi reservoirs with distribution function
$f_{\alpha\nu}(\veps) = 1/[\exp((\veps-\mu_{\alpha\nu})/k_BT_{\alpha})+1]$,
where $\mu_{\alpha\nu}=E_F+eV_{\alpha\nu}$ ($E_F$ is the Fermi energy) and $T_{\alpha}=T+\theta_\alpha$
($T$ is the background temperature). It is worthy to note that the electrochemical potential
$\mu_{\alpha\nu}$ depends on the orbital index $\nu$ that labels the bias $V_{\alpha\nu}$.
This model is valid for, e.g., a long carbon nanotube with a depleted region acting as a quasi-localized
level (the quantum dot). It has been experimentally confirmed that the orbital index is conserved
during tunneling across a highly symmetric carbon-nanotube quantum dot \cite{jar05}.
Thus, possible orbital polarizations are determined
from the imbalance $\mu_{\alpha+}\neq \mu_{\alpha-}$ \cite{lim11}.
Finally, $\theta_\alpha$ is the temperature shift applied to lead $\alpha$.

We define the {\em orbital} current as $I_o = I_{+} - I_{-}$ while the electric (charge) current
is accordingly given by $I_c = I_{+} + I_{-}$. The applied thermal difference is denoted with
$\Delta T = \theta_L - \theta_R$. With the electrochemical potential parametrization
$\mu_{\alpha} = (\mu_{\alpha+} + \mu_{\alpha-})/2$, the electric voltage bias $\Delta V$ and the orbital
bias $\Delta V_o$ become
\begin{align}
e\Delta V &= \mu_L - \mu_R \,, \\
e\Delta V_{o} &= (\mu_{L+}-\mu_{L-}) - (\mu_{R+}-\mu_{R-}) \,.
\end{align}
Notice that Ref. \cite{rej12} proposes analogous expressions for the pure spin case.

We define the orbital thermopower,
\beq
S_{o} = -\left.\frac{e\Delta V_{o}}{\Delta T}\right|_{I_c=0,I_o=0}  \,,
\edq
as the ratio between the induced orbital voltage $\Delta V_{o}$ and the applied temperature
difference $\Delta T$, in close analogy with the charge Seebeck coefficient,
\beq
S_c = \left.-\frac{e\Delta V}{\Delta T}\right|_{I_c=0,I_o=0}\,.
\edq
We emphasize that the two coefficients are calculated under the condition
that {\em both} orbital and charge currents simultaneously vanish.

In linear response, the differences $\Delta T$, $e\Delta V$, and $e\Delta V_{o}$ are small and we expand
$I_o$ and $I_c$ to first order:
\begin{align}
I_o  &= \frac{e}{h} \left[(\cali_{1+}-\cali_{1-})\frac{\Delta T}{T} 
+ (\cali_{0+}-\cali_{0-})e\Delta V
+ \frac{1}{2}(\cali_{0+}+\cali_{0-})e\Delta V_{o}\right] \,,\\
I_c &=
\frac{e}{h} \left[(\cali_{1+}+\cali_{1-})\frac{\Delta T}{T}
+ (\cali_{0+}+\cali_{0-})e\Delta V 
+ \frac{1}{2}(\cali_{0+}-\cali_{0-})e\Delta V_{o}\right] \,.
\end{align}
Here, $\cali_{n\nu}$ is the transport integral defined by
\beq
\cali_{n\nu} = \sum_{\sm}\int d\veps\, \veps^n \left(-\partial_{\veps}f_0(\veps)\right) \calt_{\nu\sm}(\veps)\,,
\edq
with $f_0(\veps) = 1/(e^{\veps/T} + 1)$ the equilibrium distribution function (we set $E_F=0$ and $k_B = 1$).

To have purely orbital currents, the charge current must vanish. This is accomplished
by the application of the electric bias
\beq
e\Delta V = -\frac{1}{2}\left(\frac{\cali_{1+}}{\cali_{0+}}+\frac{\cali_{1-}}{\cali_{0-}}\right) \frac{\Delta T}{T} \,.
\label{eq:appVO}
\edq
Therefore, the orbital Seebeck coefficient becomes
\beq
S_{o} = \frac{1}{T}\left(\frac{\cali_{1+}}{\cali_{0+}} - \frac{\cali_{1-}}{\cali_{0-}}\right) \,.
\label{eq:Sorb}
\edq
This is a general result. We expect the formation of an orbital bias $\Delta V_{o} =-eS_0\Delta T$
in the leads when the transmission depends on the orbital index,
similarly to the temperature driven generation of spin biases in junctions showing
spin-dependent scattering \cite{joh87}.

At low temperatures, it is useful to consider the Sommerfeld expansion \cite{ashcroft}.
Then,
\beq
S_{o} \xrightarrow[T\to 0]{} \frac{\pi^2 T}{3} \left(\frac{\sum_{\sm}\partial_{\veps} A_{+\sm}(E_F)}{\sum_{\sm} A_{+\sm}(E_F)} - \frac{\sum_{\sm}\partial_{\veps} A_{-\sm}(E_F)}{\sum_{\sm} A_{-\sm}(E_F)}\right) \,,
\label{eq:Sorb0}
\edq
to leading order in $T$.
This expression is a generalization of the Mott formula \cite{jon80} valid for orbital bias driven quantum systems.

For comparison, we also give the expression of the charge Seebeck coefficient:
\beq
S_c  = \frac{1}{2T}\left(\frac{\cali_{1+}}{\cali_{0+}}+\frac{\cali_{1-}}{\cali_{0-}}\right) \,,
\label{eq:Scharge}
\edq
which in the limit of $T\to 0$ becomes
\beq
S_{c} \xrightarrow[T\to 0]{} \frac{\pi^2 T}{6} \left(\frac{\sum_{\sm}\partial_{\veps} A_{+\sm}(E_F)}{\sum_{\sm} A_{+\sm}(E_F)} + \frac{\sum_{\sm}\partial_{\veps} A_{-\sm}(E_F)}{\sum_{\sm} A_{-\sm}(E_F)}\right) \,.
\label{eq:Scharge0}
\edq

Equations \eqref{eq:Sorb} and \eqref{eq:Scharge} are valid for generic nanostructures
with two orbital states. As an illustration, we now consider a quantum dot with energy levels
\beq
\varepsilon_\nu=\varepsilon_d+\nu\frac{\delta}{2}\,,
\edq
where $\delta$ is the orbital splitting induced by any symmetry breaking mechanism
such as a magnetic field along a nanotube axis \cite{cho05}
and $\varepsilon_d$ is the mean energy level measured with respect to $E_F$.
Formally, the problem is equivalent to a spin-split quantum dot with a single
energy level. However, the difference is that spin and orbital states couple differently
to an external magnetic field since their associated magnetic moments generally
differ; e.g., for a carbon-nanotube quantum dot, orbital splittings of the order
of $\delta$ are 10--20 times larger than spin splittings at a fixed magnetic field \cite{min04}.
We shall first consider noninteracting electrons and then discuss in detail
the strongly correlated case where the orbital degree of freedom plays a crucial role.

\section{Noninteracting limit}

\begin{figure}
\centering
\includegraphics[width = 0.45\textwidth]{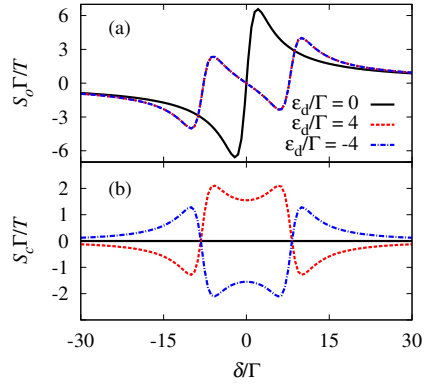}
\caption{(a) Orbital ($S_o$) and (b) charge ($S_c$) Seebeck coefficients as a function of the level splitting
$\delta$ in the noninteracting limit and temperature $T\to 0$ for various values of the level position
$\varepsilon_d$.}
\label{fig:NISC}
\end{figure}

For noninteracting electrons, the exact expression for the dot spectral weight is
\beq
A_{\nu\sm}(\veps) = \frac{1}{\pi}\frac{\Gamma}{(\veps-\veps_{\nu})^2+\Gamma^2} \,.
\edq
Using this equation in Eqs.\ \eqref{eq:Sorb0} and \eqref{eq:Scharge0}
we find the low temperature behavior of the Seebeck cofficients:
\bes
\beq
S_o \xrightarrow[T\to 0]{} \frac{2\pi^2 T}{3} \left(\frac{\veps_{d} + \delta/2}{(\veps_d+\delta/2)^2+\Gamma^2}-\frac{\veps_{d} - \delta/2}{(\veps_d-\delta/2)^2+\Gamma^2}\right) \,,\label{eq:Sorb_nint}
\edq
\beq
S_{c} \xrightarrow[T\to 0]{} \frac{\pi^2 T}{3} \left(\frac{\veps_{d} + \delta/2}{(\veps_d+\delta/2)^2+\Gamma^2} + \frac{\veps_{d} - \delta/2}{(\veps_d-\delta/2)^2+\Gamma^2}\right) \,,
\edq
\eds
which are plotted in Fig.~\ref{fig:NISC}.
We observe in Fig.~\ref{fig:NISC} (b) that when $\delta=0$ the charge thermopower $S_c$ changes sign when the dot level
$\veps_d$ lies above or below $E_F$. This is an expected behaviour due to the ability of $S_c$ to indicate
electron- or hole-like transport \cite{red07}. As $\delta$ increases, $S_c$ remains roughly constant until
the split level crosses $E_F$ and $S_c$ then changes sign. Importantly, the charge thermopower
vanishes at the particle symmetry point  ($\veps_d = 0$) regardless of the $\delta$ value.

More interestingly, the orbital thermopower $S_o$ shows distinct features, see Fig.~\ref{fig:NISC}(a).
It vanishes in both limits, $\delta \to 0$ and $\delta \gg \Gamma$.
This is expected since no orbital bias can be induced if the two orbitals are degenerate
or they lie far apart. Furthermore, the orbital thermopower 
is quite generally nonzero when particle-hole symmetry takes place at $\veps_d = 0$,
unlike $S_c$. The two Seebeck coefficients also differ when transport
is electron- or hole-like. While $S_c$ changes its sign when  $\veps_d$ is reversed with respect to $E_F$,
the orbital Seebeck coefficient is insensitive
to whether transport is dominated by electron or hole excitations
since both curves for $\veps_d = 4\Gamma$ and $\veps_d = -4\Gamma$
in Fig.~\ref{fig:NISC}(a) are identical [$S_o(\veps_d)=S_o(-\veps_d)$ in Eq.~\eqref{eq:Sorb_nint}].
In addition, for $\veps_d=0$ $S_0$ reaches an optimal
value when the spliting $\delta$ is of the order of $\Gamma$ because charge fluctuations
are maximal precisely at that level position. The optimal value shifts with $\veps_d\neq 0$
and new peaks arise due to the passage of the split level $\varepsilon_\nu$ across $\sim\pm\Gamma$
above and below the Fermi energy. This demonstrates a full tunability of the generated
orbital population with the aid of an external gate voltage.

\section{Strong coupling regime}

Consider now electron-electron interactions described by $\sum_{\nu\sigma\neq\nu'\sigma'}U n_{\nu\sigma} n_{\nu'\sigma'}$,
where $n_{\nu\sigma}$ is the occupation of the dot spin-orbital state $(\nu,\sigma)$ and
$U$ is the onsite charging energy. Using the Friedel-Langreth sum rule \cite{lan76,hew97}, 
the spectral weight $A_{\nu\sm}(\veps)$ at $\veps = E_F$ can be expressed in terms of $n_{\nu\sigma}$:
\beq
A_{\nu\sm}(E_F) = \frac{\sin^2(n_{\nu\sm}\pi)}{\pi\Gamma} \,.
\label{eq:A0}
\edq
It follows that its energy derivative takes the form \cite{rou12}
\beq
\partial_{\veps} A_{\nu\sm}(E_F) = \frac{1}{\pi \Gamma\tGamma_{\nu\sm}} \sin(2n_{\nu\sm}\pi) \sin^2(n_{\nu\sm}\pi) \,.
\label{eq:Ap0}
\edq
Here, the tunnel broadening $\tGamma_{\nu\sm} = z_{\nu\sm}\Gamma$ becomes renormalized
by the quasi-particle weight factor $z_{\nu\sm} = 1/[1-\partial_{\veps}\Re\Sigma_{\nu\sm}^r(E_F)]$,
where $\Sigma_{\nu\sm}^r$ is the retarded self-energy contribution due to interaction effects \cite{hew97}.

Combining Eqs.~\eqref{eq:A0} and \eqref{eq:Ap0}, we find the thermopowers
\bes
\begin{align}
S_o &= \frac{\pi^2 T}{3}(\cals_{+} - \cals_{-}) \,, \\
S_c &= \frac{\pi^2 T}{6}(\cals_{+} + \cals_{-}) \label{eq:SchaF} \,,
\end{align}
\label{eq:SOSC}
\eds
where
\beq
\cals_{\nu} \equiv \frac{\cali_{1\nu}}{\cali_{0\nu}} = \frac{\sum_{\sm}\left[\sin(2n_{\nu\sm}\pi)\sin^2(n_{\nu\sm}\pi)/\tGamma_{\nu\sm}\right]}{\sum_{\sm}\sin^2(n_{\nu\sm}\pi)} \,.
\label{eq:Snu}
\edq
Since our system is spin rotationally invariant, we have  $n_{\nu\sm} = n_{\nu}/2$ and $\tGamma_{\nu\sm} = \tGamma_{\nu}$. Thus, Eq.~\eqref{eq:Snu} can be further simplified:
\beq
\cals_{\nu} = \frac{1}{\tGamma_{\nu}} \sin(n_{\nu}\pi) \,,
\label{eq:Snu2}
\edq 
where $n_{\nu} = \sum_{\sm} n_{\nu\sm}$.

Equations \eqref{eq:SOSC} and \eqref{eq:Snu2} are formally exact in the strong coupling regime,
i.e., when temperature is much lower than
the characteristic Kondo temperature of the system.
Our goal is then to find the orbital occupation $n_{\nu}$, which fully determines both
the orbital and charge Seebeck coefficients. One possibility is to employ a slave-boson mean-field
theory \cite{rou12}. However, this approach neglects the orbital index in the renormalized
hybridization function, $\tGamma_{\nu} \simeq \tGamma$. This is qualitatively correct
in the limit $\delta \to 0$ but it breaks down as $\delta$ increases because $\tGamma_{\nu}$ will be renormalized differently
for $\nu=\pm$, similarly to the spin Kondo effect in the presence of ferromagnetism \cite{mar03,cho03,pas04}.
Since our main goal in the remainder of the paper is to discuss a qualitative picture of the orbital themoelectric effect
in a strongly correlated system, we prefer not to delve into complicated details and consider instead
the {\em scaled} thermopowers
\bes\label{eq:SOSC2}
\begin{align}
\tilde{S}_o &= \frac{\pi^2}{3}(\cals_{+}\tGamma_{+}-\cals_{-}\tGamma_{-}) = \frac{\pi^2}{3}(\sin(n_{+}\pi)-\sin(n_{-}\pi)) \,, \\
\tilde{S}_c &= \frac{\pi^2}{6}(\cals_{+}\tGamma_{+}+\cals_{-}\tGamma_{-}) = \frac{\pi^2}{6}(\sin(n_{+}\pi)+\sin(n_{-}\pi))\,.
\end{align}
\eds

Next, we follow two different routes for assessing $n_\nu$. First, we consider a variational approach that yields
analytical results for the Kondo temperature and the dot orbital occupation. Then, we perform
a numerical renormalization group analysis which fully takes into account Kondo fluctuations in the orbital states.

\section{Variational approach}

We consider the limit $U\to \infty$. Since
the Kondo ground state is a many-body singlet, we take the trial wave function \cite{gsc88}
\beq
|\psi_0\rangle = \left(\alpha + \sum_{k}^{k_F}\sum_{\nu,\sm}\beta_{k\nu}d_{\nu\sm}^{\dag}c_{k\nu\sm}\right)|F\rangle \,,
\label{eq:trialwf}
\edq
where $c_{k\nu\sm}^{\dag}(c_{k\nu\sm})$ ($d_{\nu\sm}^{\dag}(d_{\nu\sm})$) annihilates (creates)
a conduction (dot) electron with momentum $k$ and spin $\sm$ in a channel $\nu$
and $|F\rangle$ represents the filled Fermi sea ground state when the dot states are empty.

To calculate the variational energy of the trial wave function, we use the energy functional
\beq
E_0[|\psi_0\rangle] = \frac{\nbraket{\psi_0|\calh|\psi_0}}{\nbraket{\psi_0|\psi_0}} \,.
\label{eq:Efunc}
\edq
where the system Hamiltonian reads
\begin{equation}
\calh = \sum_{\alpha,k,\nu,\sm} \veps_{\alpha k} c_{\alpha k\nu\sm}^{\dag}c_{\alpha k\nu\sm}
+ \sum_{\nu,\sm} \veps_{\nu} d_{\nu\sm}^{\dag} d_{\nu\sm} 
+ \sum_{\alpha,k,\nu,\sm} \left(t_{\alpha} c_{\alpha k\nu\sm}^{\dag} d_{\nu\sm} + \hc\right) \,.
\label{eq:MHam}
\end{equation}
with the constraint that the dot occupation is always 1 due to the infinite charging energy limit.

On minimizing Eq. \eqref{eq:Efunc} with respect to $\alpha$ and $\beta_{k\nu}$ we find
\beq
E_0 = \sum_{k,\sm} \frac{t^2}{\veps_k - T_K} + \sum_{k,\sm} \frac{t^2}{\veps_k - T_K - \delta} \,,
\label{eq:vepsTK}
\edq
where $t=\sqrt{t_L^2+t_R^2}$. The Kondo temperature is defined as $T_K = \veps_{-} - E_0$, i.e.,
the energy difference between the lowest orbital level (we take $\delta>0$) and the ground state energy.
We transform in Eq.~\eqref{eq:vepsTK} the sums over $k$ into integrals. Hence \cite{rou12,tos12},
\beq\label{eq:TKd}
T_K(\delta) = \left\{D(D+\delta)\exp\left[\frac{\pi\veps_{-}}{2\Gamma}\right] + \frac{\delta^2}{4}\right\}^{1/2} - \frac{\delta}{2} \,,
\edq
where $D$ is the lead bandwidth.
For $\delta = 0$, we have a strongly correlated four-fold degenerate state and the resulting
Kondo state possesses SU(4) symmetry with a Kondo temperature $T_K^{\rm SU(4)}=T_K(0) = D\exp[\pi\veps_{-}/4\Gamma]$.
As $\delta$ increases orbital flip transitions become energetically costly and in the limit
$\delta\to\infty$ we recover purely spin Kondo physics characterized with a Kondo temperature
$T_K^{\rm SU(2)}=T_K(\infty) = D\exp[\pi\veps_{-}/2\Gamma]$. Due to a different numerical
factor inside the exponential, one has $T_K^{\rm SU(4)}\gg T_K^{\rm SU(2)}$, as expected \cite{bor03}.

\begin{figure}
\centering
\includegraphics[width=0.45\textwidth]{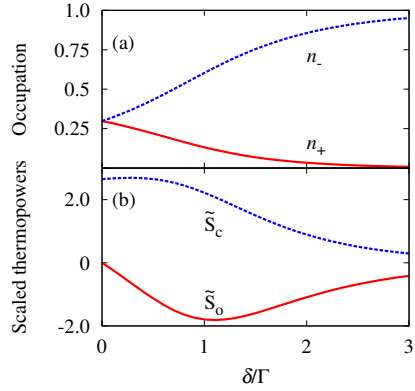}
\caption{(a) Dot occupation $n_{\nu}$ for the orbital quantum number $\nu=\pm$ as a function
of the level splitting $\delta$ obtained from a variational apprach.
(b) Scaled thermopowers (charge $\bar{S}_c$ and orbital $\bar{S}_o$)
as a function of $\delta$.
Parameters: $\veps_d/\Gamma = -4$, $D/\Gamma = 20$, $U\to\infty$
and $T\to 0$.}
\label{fig:Occupation}
\end{figure}

The average dot occupation is given by
\beq
n_{\nu} = \frac{\nbraket{\psi_0|\sum_{\sm}d_{\nu\sm}^{\dag}d_{\nu\sm}|\psi_0}}{\nbraket{\psi_0|\psi_0}}
= \frac{\sum_{k,\sm} \beta_{k\nu}^2}{\alpha^2 + \sum_{k,\nu,\sm} \beta_{k\nu}^2} \,.
\label{eq:nnu0}
\edq
The minimization procedure and the integration over the $k$-space yield:
\beq
n_{\nu}(\delta) = \frac{2\Gamma T_K\left(T_K+\delta\right)/\left(T_K+\Lambda\right)}
{\pi T_K\left(T_K+\delta\right) + 2\Gamma\left(2T_K + \delta\right)} \,,
\label{eq:nnu}
\edq
where $\Lambda=\delta$($0$) if $\nu=+$($-$). We recall that $T_K$ is a function of $\delta$, cf. Eq. \eqref{eq:TKd}.
When $\delta=0$, the occupation is the same for both orbital levels:
\beq
n_{\nu} \xrightarrow[\delta = 0]{} \frac{2\Gamma}{\pi T_K^{\rm SU(4)} + 4\Gamma} \,.
\label{eq:nnud0}
\edq
As $\delta$ increases, the orbital $\nu=+$ becomes less populated due to the level splitting,
as depicted in Fig.\ \ref{fig:Occupation}(a) with solid lines. In the SU(2) Kondo limit ($\delta\to\infty$), Eq.~\eqref{eq:nnu} gives
\beq
n_{\nu} \xrightarrow[\delta\to\infty]{}
\begin{cases}
\pi T_K^{\rm SU(2)} & \text{for $\nu = +$} \,, \\
2\Gamma/\left(\pi T_K^{\rm SU(2)} + 2\Gamma\right) & \text{for $\nu = -$} \,.
\end{cases}
\edq
In general, the SU(2) Kondo temperature is much smaller then the hybridization width,
$T_K^{\rm SU(2)}\ll\Gamma$. Therefore, $n_-=1$ and $n_+=0$ to a good
extent [see Fig.\ \ref{fig:Occupation}(a)] and we recover the $1/2$ value of the population per spin
obtained at very low temperatures \cite{hew97}.

\begin{figure}
\centering
\includegraphics[width=0.45\textwidth]{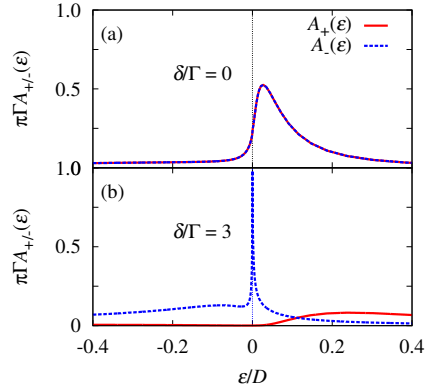}
\caption{Numerical renormalization group calculation
of the dot spectral weight $A_{\nu}$ as a function of energy $\veps$ for the two
orbital states $\mu=\pm$. (a) SU(4) Kondo resonance clearly develops for level splitting $\delta=0$
(the spectral weights for both orbitals coincide).
(b) SU(2) Kondo resonance forms when $\delta$ is tuned beyond the crossover point between
the two Kondo states in which case the contribution from the $\nu=+$ channel to the Kondo resonance
is negligible. Parameters: $\veps_d/\Gamma = -4$, $D/\Gamma = 20$, $U/\Gamma = 200$ and $T\to 0$.
The vertical dotted line is a guide to the eye.}
\label{fig:LDOS}
\end{figure}

Clearly, the orbital level occupations differ depending on the Kondo state symmetry.
As a consequence, the thermopowers (orbital and charge) will be significantly altered
as a function of the level splitting $\delta$.  Furthermore,
for a system with SU(2) symmetry the Kondo resonance develops at the Fermi level $E_F$,
see Fig.~\ref{fig:LDOS}(a). (We below discuss the numerical method that generates Fig.~\ref{fig:LDOS}).
Therefore, the charge thermopower $S_c$ will attain an exceedingly small value at low temperatures since
the derivative of the spectral weights $\partial_{\veps} A_{\nu\veps}(\veps)$ vanishes at $E_F$.
The dashed line in Fig.~\ref{fig:Occupation}(b) at $\delta\gg\Gamma$ precisely reflects this property.
On the other hand, for a system with SU(4) symmetry the Kondo resonance develops at $\veps \approx T_K^{SU(4)}$,
as shown in Fig.~\ref{fig:LDOS}(b).
This is a crucial difference with the SU(2) case since $\partial_{\veps} A_{\nu\veps}(E_F) \ne 0$
and $\tilde{S_c}$ then reaches a finite value at $\delta = 0$.

More interestingly, the orbital Seebeck coefficient $\tilde{S}_o$ reaches a maximum (in absolute value)
at intermediate values of the level splitting, see Fig.~\ref{fig:Occupation}(b). At $\delta=0$, $\tilde{S}_o$
vanishes because $n_+=n_-$.
For $\delta\gg\Gamma$, $\tilde{S}_o$ tends to zero for the same reason that
the charge thermopower decreases---the Kondo resonance remains pinned at $E_F$.
Then, an extremum must arise for a nonzero value of $\delta$.
We find that a maximal orbital bias is generated when the splitting is of the order of
$T_K^{\rm SU(4)}$. Since this energy scale is precisely of the order of the
level broadening, our results can be understood in terms of
a resonance which behaves effectively as a noninteracting
system with renormalized  parameters. This picture is valid
in the low temperature regime where
Kondo correlations simultaneously quench spin and charge fluctuations \cite{hew97}.

\section{Numerical results}

Our previous results were restricted to $U\to\infty$ case. We now consider large (but finite)
charging energies using a numerical renormalization group (NRG) formalism.

In the Lehmann representation, the dot spectral weight takes the form
\beq\label{eq:ANRG}
A_{\nu\sm}(\veps) 
= \frac{1}{\calz f_0(\veps)}\sum_{p,q} e^{-E_p/T} |\langle p|d_{\nu\sm}^{\dag}|q\rangle|^2 \delta\left(\veps-(E_p-E_q)\right) \,,
\edq
where $\calz = \sum_p e^{-E_p/T}$ is the partition function
and $E_p$, $E_q$ are many-body eigenenergies calculated within NRG \cite{bul08}.
We use Eq.~\eqref{eq:ANRG} to calculate the dot local densities of states
shown in Fig.~\ref{fig:LDOS}.

\begin{figure}
\centering
\includegraphics[width=0.45\textwidth]{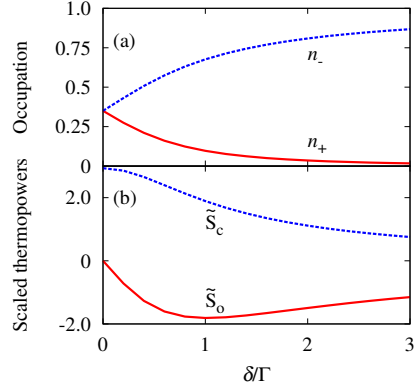}
\caption{(a) NRG dot occupation $n_{\nu}$ for the orbital quantum number $\nu=\pm$ as a function
of the level splitting $\delta$. (b) NRG scaled thermopowers (charge $\tilde{S}_c$ and orbital $\tilde{S}_o$)
as a function of $\delta$.
Parameters: $\veps_d/\Gamma = -4$, $D/\Gamma = 20$,
$U/\Gamma = 200$ and $T\to 0$.}
\label{fig:Occupation2}
\end{figure}

The orbital occupation is readily obtained from Eq.~\eqref{eq:ANRG} as
\beq\label{eq:nNRG}
n_{\nu}=\sum_\sm\int d\veps \, A_{\nu\sm}(\veps) f_0(\veps)\,.
\edq
In Fig.~\ref{fig:Occupation2}(a), we depict $n_{\nu}$
for $U=200\Gamma$ as a function of $\delta$.
For vanishingly small level splittings, the occupations are equal, $n_{+} = n_{-}$, as expected.
Importantly, their exact values are smaller than $1/2$. This can be understood with the aid of
Eq.~\eqref{eq:nnud0}. 
Unlike the exponentially small SU(2) Kondo temperature $T_K(\infty)$, 
the higher SU(4) Kondo temperature is $T_K^{\rm SU(4)}\simeq 0.864 \Gamma $ for the parameters
used in Fig.~\ref{fig:Occupation2}. Therefore, its contribution cannot be neglected in the denominator
of Eq.~\eqref{eq:nnud0}. This is a crucial difference with the SU(2) case.
In addition, when $\delta$ increases $n_{+}$ tends to vanish since the level $\veps_{+}$
is pushed up and its occupation is energetically hindered. At the same time,
$n_{-}$ shows the opposite behavior. 

Figure~\ref{fig:Occupation2}(b) shows the scaled thermopowers obtained from our NRG calculations.
Our results strongly resemble those obtained with the variational approach,
cf. Fig.~\ref{fig:Occupation}(b). This confirms our previously discussed picture of the orbital thermopower minimum signaling
the transition from SU(4) to SU(2) Kondo physics as the level splitting is increased.
Notice that here we have analyzed scaled Seebeck coefficients
since they are easier to understand (they depend on the occupation only, see Eqs.~\eqref{eq:SOSC2}).
We do not expect
qualitative changes if the exact $S$ were calculated using, e.g., the methods
discussed in Refs.~\cite{rej12,yos09,ser09}.


\section{Conclusions}

We have investigated the formation of orbital accumulations in systems with spin and pseudosin
degrees of freedom under the influence of externally applied temperature differences.
We have defined the orbital Seebeck coefficient from an open-circuit pure orbital bias.
We have found that orbital thermopower is really sensitive to changes in level splitting
fields possibly present in the system. Thus, we propose to use the occurrence of
orbital thermopower peaks as the 'smoking gun' of the transition between
Kondo states with distinct symmetry types.

The presence of orbital polarizations could be experimentally detected using
the different coupling of circularly polarized light to the unequal population
of electronic orbital states \cite{ber05}. An alternative scheme might measure
the magnetization response using ultrasmall magnetometers \cite{lev90}.
Further work is thus needed to test the effects discussed in this paper.

\ack This work was supported by MINECO Grants No. FIS2011-2352 and CSD2007-00042 (CPAN).

\end{document}